\documentclass[preprint,12pt]{elsarticle}
\usepackage{amssymb}

\def\J{J/\Psi}

 \def\gsim{\mathrel{\rlap{\lower4pt\hbox{\hskip1pt$\sim$}}
 \raise1pt\hbox{$>$}}}

 \newcommand\la{\langle}
 \newcommand\ra{\rangle}
 \newcommand\beq{\begin{equation}}
 
 \newcommand\eeq{\end{equation}}
 \newcommand\beqn{\begin{eqnarray}}
 \newcommand\eeqn{\end{eqnarray}}

\def\fm{\,\mbox{fm}}
\def\GeV{\,\mbox{GeV}}
\def\TeV{\,\mbox{TeV}}
\def\lsim{\mathrel{\rlap{\lower4pt\hbox{\hskip1pt$\sim$}}
    \raise1pt\hbox{$<$}}}         
\def\gsim{\mathrel{\rlap{\lower4pt\hbox{\hskip1pt$\sim$}}
    \raise1pt\hbox{$>$}}}         

\def\fm{\,\mbox{fm}}
\def\GeV{\,\mbox{GeV}}

\def\beq{\begin{equation}}
\def\eeq{\end{equation}}

\def\beqy{\begin{eqnarray}}
\def\eeqy{\end{eqnarray}}

\begin{document}

\begin{frontmatter}

\title{Puzzles of $J/\Psi$ production off nuclei}

\author{B. Z. Kopeliovich}

\address{Departamento de F\'{\i}sica, 
Universidad T\'ecnica Federico Santa Mar\'{\i}a, and
\\
Instituto de Estudios Avanzados en Ciencias e Ingenier\'{\i}a, and\\
Centro Cient\'ifico-Tecnol\'ogico de Valpara\'iso;\\
Casilla 110-V, Valpara\'iso, Chile}

\begin{abstract}

Nuclear effects for $\J$ production in $pA$ collisions are controlled by the coherence and color transparency effects. Color transparency onsets when the time of formation of the charmonium wave function becomes longer than the inter-nucleon spacing. In this energy regime the effective break-up cross section for a $\bar cc$ dipole depends on energy and nuclear path length, and agrees well with data from fixed target experiments, both in magnitude and energy dependence. At higher energies of RHIC and LHC coherence in $\bar cc$ pair production leads to charm  quark shadowing
which is a complement to the high twist break up cross section. These two effects explain well with no adjusted parameters the magnitude and rapidity dependence of nuclear suppression of $\J$ observed at RHIC in $dAu$ collisions, while the contribution of leading twist gluon shadowing is found to be vanishingly small.
A novel mechanism of double color filtering for $\bar cc$ dipoles makes nuclei significantly more transparent in $AA$ compared to $pA$ collisions. This is one of the mechanisms which make impossible a model independent "data driven" extrapolation from $pA$ to $AA$. This effect also explains the enhancement of nuclear suppression observed at forward rapidities  in $AA$ collisions at RHIC, what hardly can be related to the produced dense medium. $\J$ is found to be a clean and sensitive tool measuring the transport coefficient characterizing the dense matter created  in $AA$ collisions.
RHIC data for $p_T$ dependence of $\J$ production in nuclear collisions are well explained with the low value of the transport coefficient $\hat q_0<0.5\GeV^2/\fm$.

 \end{abstract}



\end{frontmatter}

\section{Time scales and different regimes for $\J$ attenuation in nuclei}

 The widely used model for high twist nuclear effects is based on unjustified assumptions:
 (i) $J/\Psi $ (a c-cbar dipole) is always created momentarily inside the nucleus; (ii)
the produced $\bar cc$ dipole attenuates in the nucleus with a break-up cross section $\sigma_{abs}$, which is assumed to be universal for all nuclei, independent of energy and $x_F$, and is fitted to data.

Let us start up with examining the latter assumption (ii), assuming for the moment that the former one (i) is true.

\subsection{$\J$ formation, color transparency, break-up cross section}

A  $\bar cc$   dipole is produced with a small separation $r_{\bar cc}\sim1/m_c\approx0.1\fm$.
Then it evolves into a $\J$ whose mean size is quite larger, $r_{\J}\approx0.4\fm$. Correspondingly, the absorption cross section, which scales as $r^2$, increases by an order of magnitude.
The expansion time is given by the uncertainty principle,
\beq
t_f=\frac{2E_{\J}}{M_{\Psi'}^2-M_{\J}^2}.
\label{20}
\eeq
Indeed, the produced $\bar cc$ dipole has a certain size, but no certain mass, and it takes time to resolve between the $\J$ and the nearest radial excitation $\Psi'$. There are in fact several time scales controlling the expansion process, the one given by Eq.~(\ref{20}) is the longest.

A low energy dipole quickly expands to $\J$, while at high energy Lorentz time dilation freezes the initial small size for the time of propagation through the nucleus. So the nuclear medium becomes more transparent with rising energy, i.e. the effective break-up cross section decreases.

To quantify this effect, let us consider a simplified equation based on the uncertainty relation describing the transverse  expansion of a $\bar cc$ dipole moving with energy $E_{\bar cc}$,
\beq
\frac{dr_T}{dt}=\frac{4p_T}{E_{\bar cc}}.
\label{40}
\eeq
Applying the uncertainty relation $p_T\sim1/r_T$, we get a solution
\beq
r_T^2(t)=\frac{8t}{E_{\bar cc}}+\frac{\delta}{m_c^2}.
\label{60}
\eeq
Here the initial separation squared of the $\bar cc$ dipole is fixed at the value $\la r_T^2\ra\sim \delta/m_c^2$, which deserves a discussion. At high energies the amplitude factorizes into the light-cone size distribution amplitude of $\bar cc$ fluctuations in a gluon, given by the modified Bessel function $K_0(m_c r_T)$,
and the amplitude of $\bar cc$ dipole  interaction with the target nucleon \cite{kz91}. However, at low energies, when the time of charm production becomes as short as the proton radius, such factorization breaks down and the size distribution is poorly known. So in Eq.~(\ref{60})  $\delta\sim1$, but not known more accurately (see more detailes in \cite{kz91}). To evaluate the theoretical uncertainty we will try $\delta=0.5,\ 1.0,\ 2.0$.

 Notice that the solution Eq.~(\ref{60}) is valid only for $t\lsim t_f$, when the quarks can be treated as free particles. At longer times the $c$-$\bar c$ interaction becomes important and affects the expansion process. A rigorous solution based on the path-integral technique is known \cite{kz91,kst2}, but is more complicated. Here, for the sake of simplicity,  we rely on the solution Eq.~(\ref{60}) assuming that $t_f\gsim R_A$,
i.e. $E_{\bar cc}\gsim {1\over2}R_A(M_{\Psi'}^2-M_{\J}^2)$.

Due to color transparency \cite{zkl} 
the dipole cross section in the small-$r_T$ approximation has the form $\sigma_{abs}=Cr_T^2$, where the factor $C(E_{\bar cc})$ depends on dipole energy $E_{\bar cc}$ in the target rest frame. 
We can calculate the mean break-up cross section for a dipole of energy $E_{\bar cc}$ propagating and expanding along a path length $L$ in a medium with a constant density,
\beq
\bar\sigma_{abs}(L,E_{\bar cc})={1\over L}\int\limits_0^L dl\,\sigma_{abs}(l) = C(E_{\bar cc})\,\left(\frac{4L}{E_{\bar cc}}+
\frac{\delta}{m_c^2}\right).
\label{80}
\eeq
We see that the mean break-up cross section is not a constant, as usually assumed, but rises with path length $L$ and decreases with energy.
The factor $C(E_{\bar cc})$ was calculated in \cite{broad}. For example, at $x_F=0$, and the energies of the experiments NA60 at CERN SPS and E866 at Fermilab, $C(E_{\bar cc})=2.89$ and  $3.14$,  respectively.

\begin{figure}[htb]
\begin{center}
\includegraphics[width=6cm]{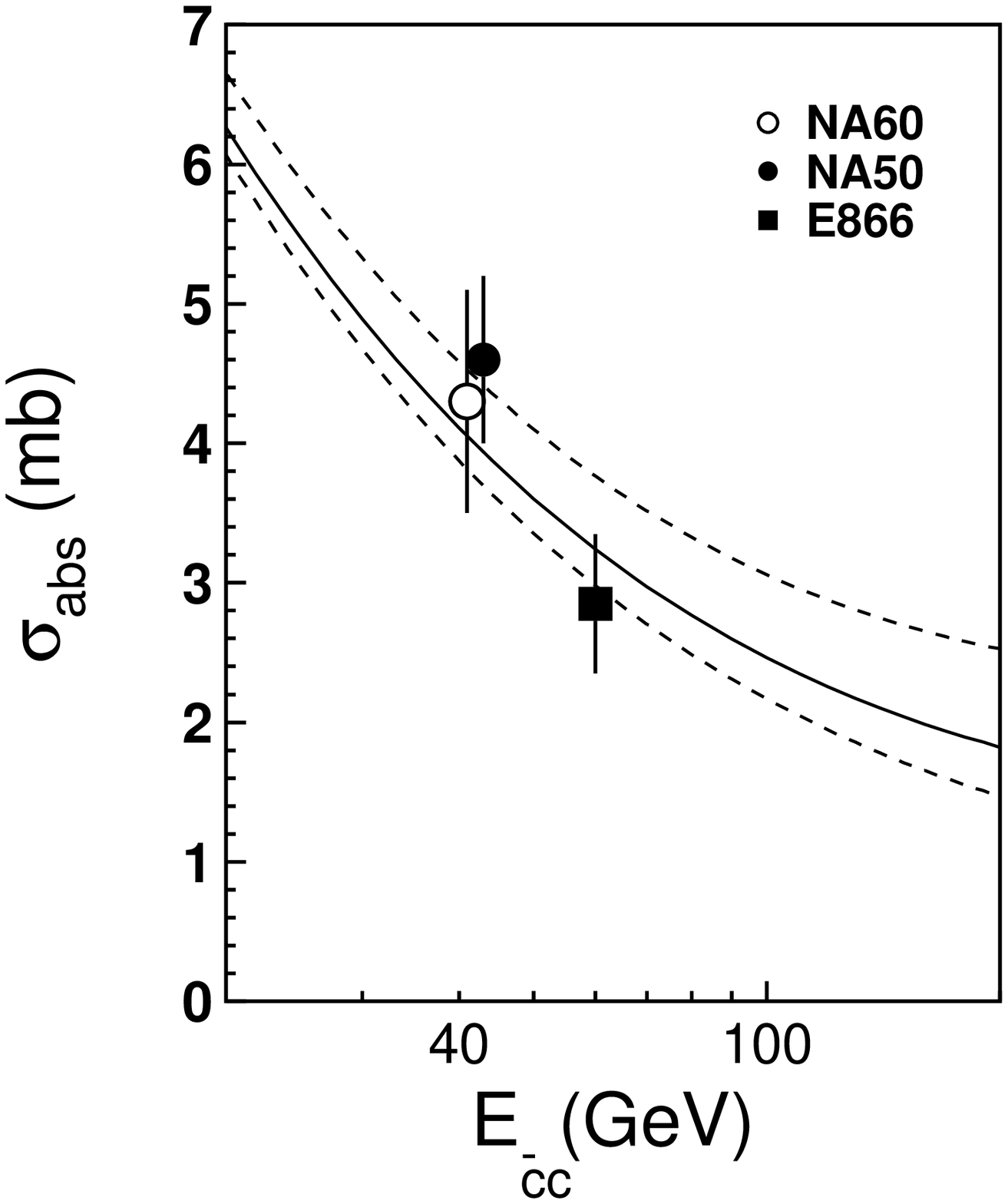}
\hspace{10mm}
\includegraphics[width=6cm]{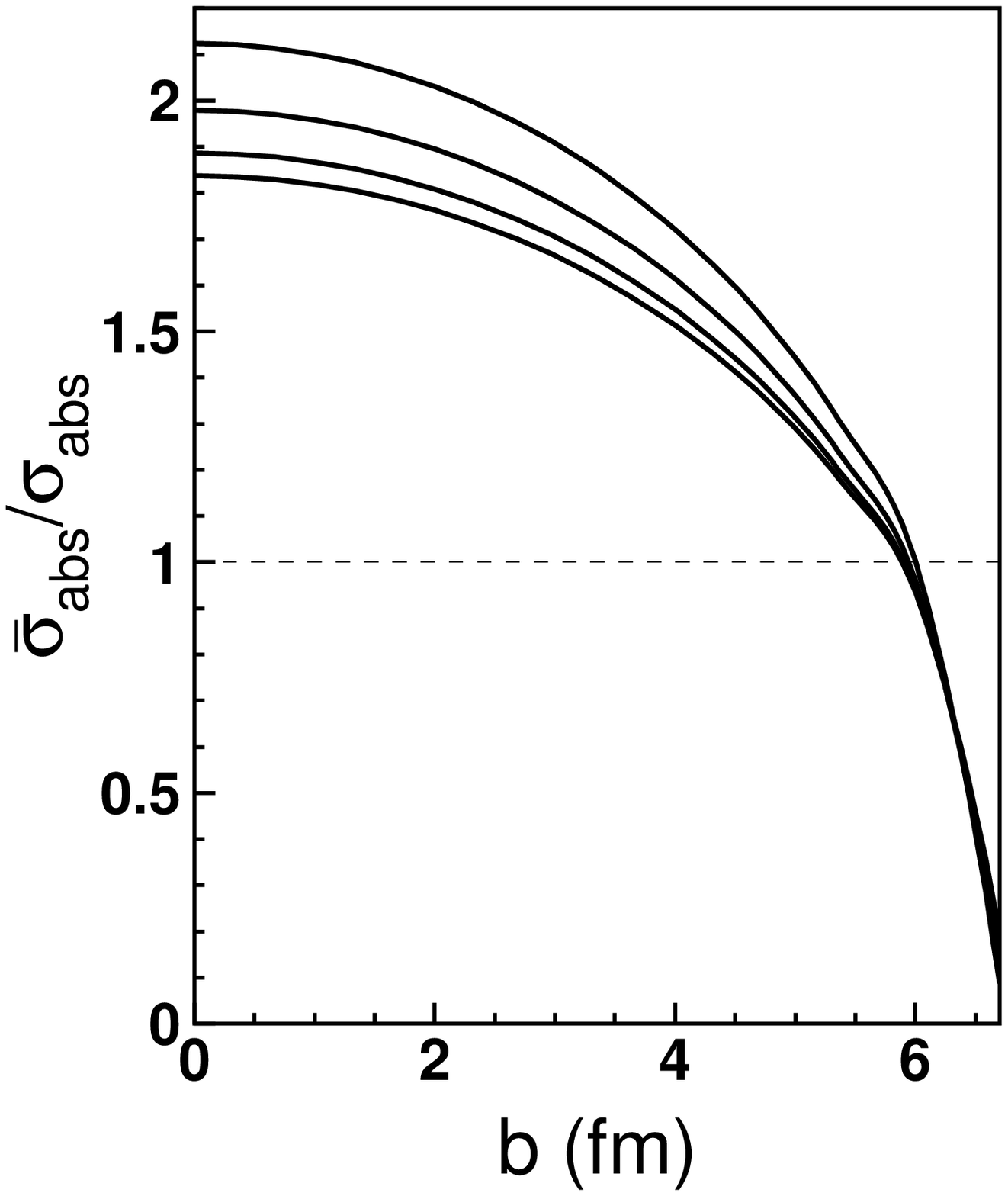}
\end{center}
\caption{\label{sig_abs} {\it Left:} the break-up cross section fitted with expression (\ref{90}) to the results of calculation with the absorption cross section Eq.~(\ref{80}) and $\delta=1$ (solid curve), as function of energy. The bottom and top dashed curves show the theoretical uncertainty corresponding to variation of $\delta=1/2,\ 2$ respectively.
Data are from fixed target experiments \cite{na60,e866}.
{\it Right:} ratio of the effective cross section Eq.~(\ref{80}) to the conventional mean one fitted 
with Eq.~(\ref{90}), as function of $b$ at different energies of $pA$ collision $E_{lab}=158,\ 400,\ 800,\ 1200\GeV$ (from top to bottom).}
 \end{figure}

Usually the nuclear ratio is evaluated with an oversimplified model assuming that $\J$ attenuates with a constant cross section $\sigma_{abs}$ on the way out of the nucleus, 
\beq
R_{pA}=\frac{1}{A\sigma_{abs}}\int d^2b\,\left[1-e^{-\sigma_{abs}T_A(b)}\right],
\label{90}
\eeq
where $\sigma_{abs}$ is treated as an unknown  parameter fitted to data. As far, as we predicted the mean break-up cross section, Eq.~(\ref{80}), we can calculate $R_{pA}$ and comparing with Eq.~(\ref{90}) extract $\sigma_{abs}$. The result is plotted as function of energy  in the left panel of Fig.~\ref{sig_abs}.
 This calculation was done with $\delta=1$ in Eq.~(\ref{80}). The theoretical uncertainty is demonstrated by two dashed curves calculated with $\delta=1/2$ (bottom) and $\delta=2$ (top).
Though with some uncertainty, our results explain well both the magnitude of $\sigma_{abs}$ and its decreasing energy dependence \cite{na60}.

Notice, that the extrapolation of $\sigma_{eff}$  up to the energies of RHIC, should be done with precautions. We remind that in this section we made an assumption about a short production time of a $\bar cc$ pair, which certainly breaks down at high energies.

Since the effective break-up cross section rises with path length, it should be larger in central than in peripheral $pA$ collisions. Indeed, in the right panel of Fig.~\ref{sig_abs} we plotted the ratio of the $L$-dependent mean break-up cross section Eq.~(\ref{80}) to the one adjusted to the total $\J$ cross section with Eq.~(\ref{90}). The mean break-up cross section significantly exceeds the fitted one in central $pA$ collisions, and underestimate it on the periphery.
For this reason an extrapolation of nuclear effects from $pA$ to $AA$ with a constant break-up cross section $\sigma_{abs}$ cannot be accurate.

\subsection{How long does it take to produce charm?}

Although the proper time of charm production is short, $t_c^*\sim 1/2m_c$, in the rest frame of the nucleus, this time linearly rises with energy,
\beq
t_c\sim\frac{2E}{M_{\J}^2}.
\label{120}
\eeq
Thus, if the energy of the produced $\J$ is sufficiently high, $E\gsim 25(\GeV)\times L(\fm)$, the effects of coherence become significant. This is a high twist shadowing in the process of a $\bar cc$ pair production by a projectile gluon. The $\bar cc$ is produced coherently in multiple interactions of the projectile gluon and the charm quarks with target nucleons. 

A good explanatory example is photoproduction of vector mesons on nuclei. While at low energies the vector meson is photo-produced inside the nucleus and then attenuates through a half of the nuclear thickness, at high energies the vector  meson appears as a Fock state of the incoming photon long prior the interaction and propagates through the whole nucleus. Therefore one expects a significantly stronger nuclear suppression at $t_c\gg R_A$, than at $t_c\ll R_A$ \cite{kz91,hk-rho}. Data for photoproduction of $\J$ \cite{benhar} and $\rho$ mesons \cite{hermes} nicely confirmed this prediction.

Since the production amplitude is convoluted with the charmonium wave function, one can assume with a good accuracy an equal sharing of the total longitudinal momentum between $c$ and $\bar c$. Then, in the small-$r_T$ approximation the amplitude of $\bar cc$ production at the point with impact parameter $b$ and longitudinal coordinate $z$ inside the nucleus, averaged over the dipole size reads \cite{kth-psi},
\beqn
&&\int d^2r_T\,W_{\bar cc}(r_T)
\exp\left[-{1\over2}C(E_{\bar cc})\, r_T^2\, \left({7\over16}T_-(b,z)
+T_+(b,z)\right)\right]\nonumber\\
&=&
\left[1+{1\over2}C(E_{\bar cc})\,\la r_T^2\ra\, \left({7\over16}T_-(b,z)
+T_+(b,z)\right)\right]^{-1}.
\label{140}
\eeqn
Here $T_-(b,z)=\int_{-\infty}^z dz'\rho_A(b,z')$;~ $T_+(b,z)=T_A(b)-T_-(b,z)$, and $T_A(b)=T_-(b,\infty)$. Although the size distribution $W_{\bar cc}(r_T)$ of produced dipoles has a complicated form, we assume for the sake of simplicity (more accurate calculations will be published elsewhere) that it has a gaussian shape,
$W_{\bar cc}(r_T)\propto e^{-r_T^2/\la r_T^2\ra}$, with the mean value $\la r_T^2\ra$, which we estimated at $\la r_T^2\ra=6/m_c^2=0.1\fm^2$.
Notice that due to color transparency the nuclear medium is more transparent than is expected in the Glauber model. Moreover, the amplitude Eq.~(\ref{140}) does not decrease exponentially with nuclear thickness, but as a power.

Integrating the amplitude Eq.~(\ref{140}) squared over coordinates of the production point, one arrives at the nuclear ratio, which has the form,
\beq
R_{pA}={1\over A}\int d^2b\,\frac{T_A(b)}
{\left[1+{1\over2}C(E_{\bar cc})\,\la r_T^2\ra\,T_A(b)\right]
\left[1+{7\over32}C(E_{\bar cc})\,\la r_T^2\ra\,T_A(b)\right]}
\label{160}
\eeq
At this point we can partially improve the small-$r_T$ approximation in (\ref{160}) replacing
$C(E_{\bar cc})\,\la r_T^2\ra\Rightarrow \sigma_{\bar qq}(r_T^2=\la r_T^2\ra)$, where
the dipole cross section has a saturated shape \cite{kst2}, and is somewhat smaller than
$C(E_{\bar cc})\,r_T^2$ at large $r_T$. 

\begin{figure}[htb]
\begin{center}
\includegraphics[width=6cm]{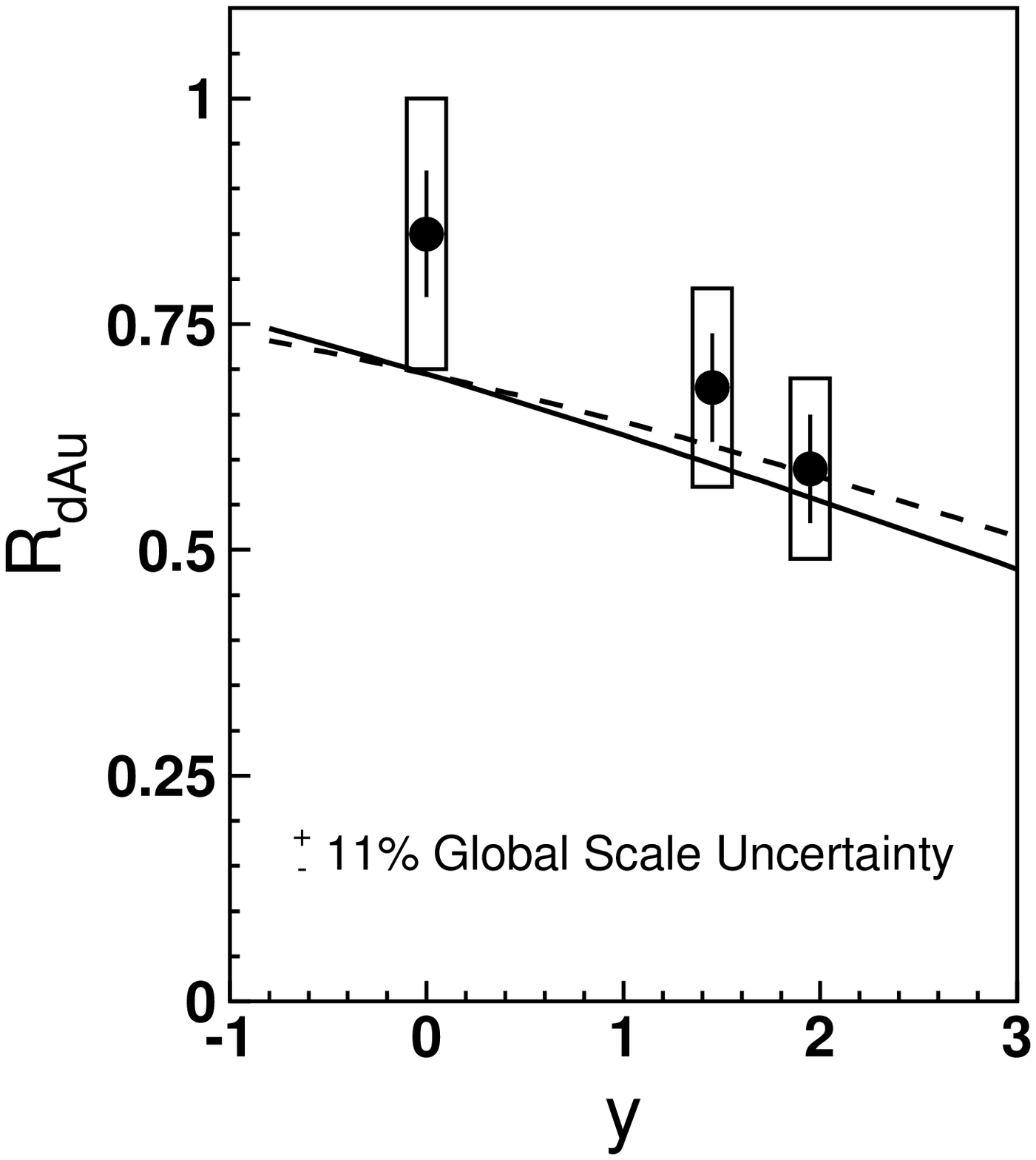}
\hspace{10mm}
\includegraphics[width=6cm]{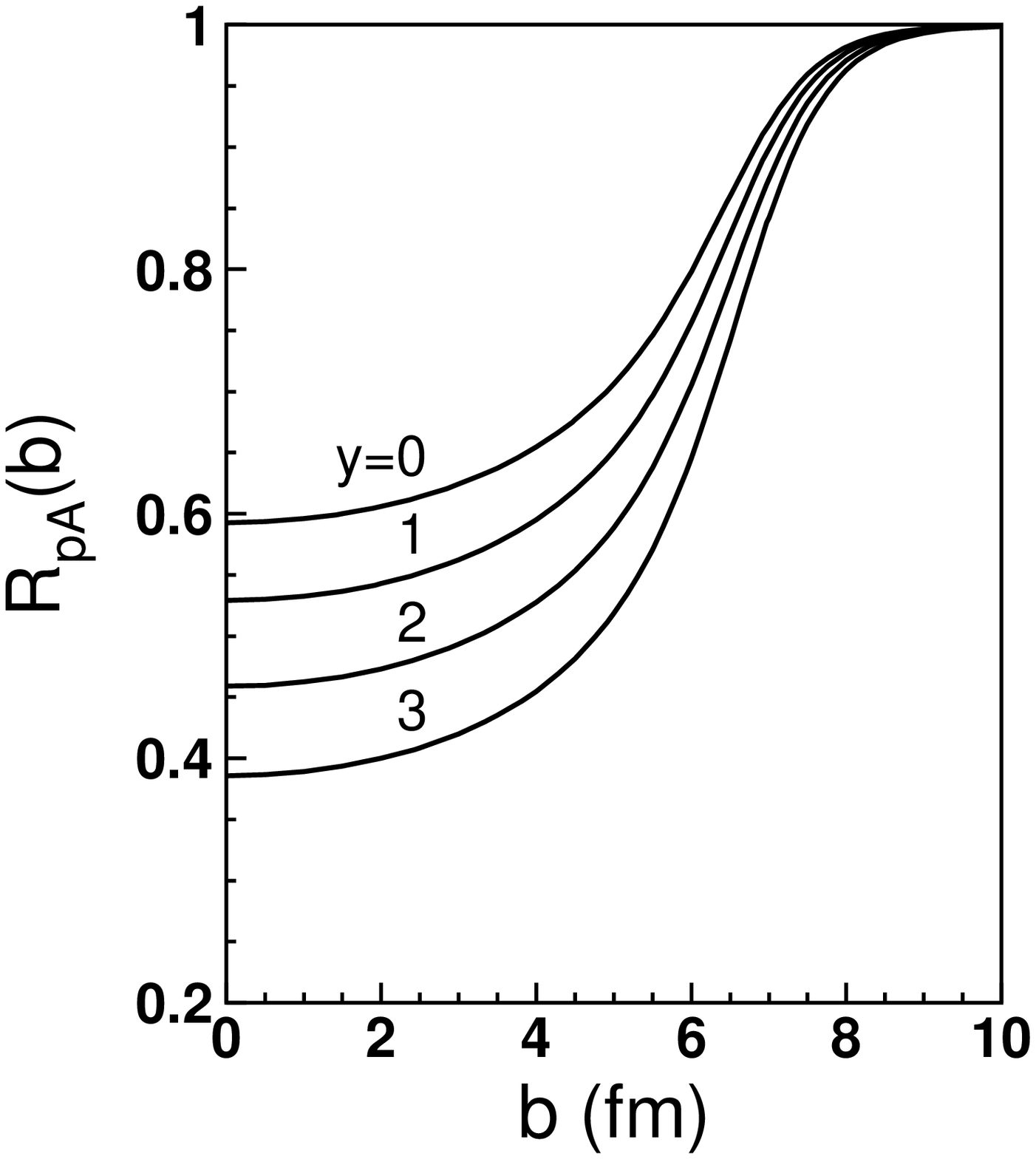}
\end{center}
\caption{\label{d-A} {\it Left:} dashed curve presents nuclear suppression of $\J$ as function of rapidity in $pA$ collisions calculated with Eq.~(\ref{160}). Solid curve is corrected for gluon shadowing. Data are for $dAu$ collisions at $\sqrt{s}=200\GeV$ \cite{phenix-dA}.
{\it Right:} $b$-dependence of the nuclear ratios for $\J$ produced with rapidities $y=0,\ 1,\ 2,\ 3$ in $pAu$ collisions at $\sqrt{s}=200\GeV$.}
 \end{figure}

With Eq.~(\ref{160}) we calculated the nuclear ratio $R_{A/p}(y)$ at $\sqrt{s}=200\GeV$, as function of  $\J$ rapidity $y$ in the c.m. of the collision, and its energy in the nuclear rest frame, $E_{\bar cc} =(\sqrt{s}/2m_N) \sqrt{M_{\J}^2+p_T^2}\,
e^{-y}$. The results are depicted in Fig.~\ref{d-A} together with data from the PHENIX experiment \cite{phenix-dA}. 
We see that the steep rise of the break-up cross section $\sigma_{\bar cc}(r_T,E_{\bar cc})$ with energy (it triples from $y=0$ to $y=2$) well explains the observed rapidity dependence of nuclear suppression. We should not continue our calculations far to negative rapidities, since the regime of long coherence length
breaks down there. Besides, additional mechanisms, which cause a nuclear enhancement at negative rapidities, must be added. 

In the right panel of Fig.~\ref{d-A} we also present the impact parameter dependence of nuclear suppression for $\J$ produced in proton-gold collisions at RHIC with different rapidities.
As expected, the strongest dependence on rapidity comes from most central collisions.

\section{Gluon shadowing}

First of all, one should evaluate the kinematic condition for gluon shadowing, $t_c^{\bar ccg}\gsim R_A$, where $t_c^{\bar ccg}$ is the coherence time, or the lifetime of a $\bar ccg$ fluctuation in a gluon.
This time can be related to the Ioffe time, as it was carefully calculated in \cite{krt2},
\beq
t_c^{\bar ccg}=\frac{P_g}{xm_N},
\label{220}
\eeq
where the factor $P_g=0.1$ was evaluated in \cite{krt2} and found to be scale-independent.
Its smallness is caused by the large intrinsic transverse momenta of gluons in hadrons, supported by numerous evidences in data \cite{spots}.
Thus, shadowing for gluons onsets at smaller $x\lsim 0.01$, than for quarks.

In $\J$ production at large $x_1$ one should redefine $x_2\Rightarrow \tilde x_2= x_2/(1-x_1)$ \cite{brodsky}. Then the smallest $\tilde x_2=M_{\J}^2/sx_1(1-x_1)$ is reached at $x_1=1/2$  
and equals to $\tilde x_2(min)=0.025$ at at the energy of the E866 experiment, $\sqrt{s}=40\GeV$.
This value is too large, so gluon shadowing has no contribution to the nuclear effects  for $\J$ production observed  in the E866 experiment , as well as in other fixed target experiments 
\cite{brahms}.

Even at the energy $\sqrt{s}=200\GeV$, the values of $x_2$ are too large for gluon shadowing  within the measured kinematics,  $\la x_2\ra=e^{-y}\,\sqrt{2M_{\J}^2+\la p_T^2 \ra}\biggl/\sqrt{s}$, where we use the $\bar cc$ invariant mass distribution predicted by the color singlet model \cite{kps-psi}.  With the measured $\la p_T^2\ra=4\GeV^2$ the value of $x_2$ ranges from $0.024$ to $0.0033$ within the measured rapidity interval $0<y<2$.  We relied upon the results of the NLO analysis  \cite{DS} of DIS data, which suggest a very weak gluon shadowing, in a good agreement with theoretical predictions \cite{kst2}. The nuclear ratio corrected for gluons shadowing at $Q^2=10\GeV^2$ \cite{DS}, 
is depicted in Fig.~\ref{d-A} by solid curve. We see that the effect of gluon shadowing is indeed vanishingly small. Even at the energy of LHC, $\sqrt{s}=5.5\TeV$ and $y=0$ gluon shadowing according to \cite{DS,kst2} is extremely small, only $3\%$ ($x_2=5.5\times10^{-3}$), and will be neglected in what follows.

Notice that our explanation of the RHIC data is quite different from the description presented in \cite{phenix-dA}. First of all, the charm quark shadowing was completely missed, and the naive
formula (\ref{90}) with a fitted break-up cross section $\sigma_{abs}$ was used. As we demonstrated, this formula is quite incorrect, especially for $b$-dependence, even at low energies, where the approximation of short coherence length is reasonable. At the high energies of RHIC and LHC, such a formula is plain wrong.

Further, the rapidity dependence of the nuclear ratio was prescribed in \cite{phenix-dA} entirely to gluon shadowing, which had a rather large magnitude. Strangely, the authors referred to the same analysis \cite{DS} as is used here, but they came up with a much stronger shadowing. Instead of the gluon shadowing fitted in \cite{DS} to data, they picked up a version called nDSg. However, the authors of \cite{DS} warned that this version should not be used as a gluon PDF, since it "should be considered only as a mean to study variations on the gluon nuclear distribution". For that purpose gluon shadowing was enforced to be large at small $x$ contradicting data. 

A similar procedure was used in \cite{eks}, although for a different reason. The magnitude of gluon shadowing was fixed "by hands" at a large value at small $x$, otherwise the LO analysis in \cite{eks} would not have had any solution for gluon shadowing at small $x$. Because of this ad hoc input, the EKS shadowing is similar in magnitude, and is as reliable, as the nDSg.

A strong gluon shadowing was reported recently in \cite{eps08}. Besides DIS data, this analysis includes data on hadron production in $dA$ collisions at forward rapidities. Interpretation of this data is still controversial \cite{brahms}, and this attempt to explain the observed  nuclear effects entirely by coherence effects led to a gluon shadowing which significantly violates the unitarity bound \cite{bound}. 

\section{Nontrivial transition from $pA$ to $AA$}\label{double-filtering}

At fist glance one might think that transition from nuclear effects in $pA$ to $AA$ collisions is straightforward: $R_{AA}(\vec b,\vec \tau)=R_{pA}(\vec\tau)\times R_{pA}(\vec b-\vec\tau)$, where
nuclei collide with impact parameter $\vec b$ and $\J$ is produced at impact parameter $\vec\tau$. Indeed, such a "data driven" procedure was used in \cite{phenix-dA,cassagnac} to predict the cold nuclear matter effects in nuclear collisions basing on measurements of $b$-dependence of nuclear suppression in $pA$. 

\subsection{Double-color-filtering}

The $pA$ to $AA$ transition, however, is not that simple. We illustrate this on the following example. If a $\bar cc$ dipole of transverse separation $r_T$ propagates through a slice of nuclear medium of thickness $T_A$,
its survival probability is $S_{pA}(r_T)=\exp(-C\,r_T^2\,T_A)$. Integrating over $r_T$ with the size distribution function $W(r_T)\propto\exp[-r_T^2/\la r_T^2\ra]$ leads to (compare with (\ref{140})),
\beq
S_{pA}=\frac{1}{1+C\la r_T^2\ra\,T_A}.
\label{240}
\eeq
In the case of a central $AA$ collision, according to the above recipe one should expect $S_{AA}(b)=
S_{pA}^2$. 

Let us, however repeat the above averaging over dipole size of  $S_{AA}(r_T)=S_{pA}^2(r_T)=
\exp(-2\times Cr_T^2 T_A)$. Factor 2 is here because the dipole attenuates simultaneously through both nuclei. The result of averaging over $r_T$ (left) should be compared with the conventional recipe (right),
\beq
S_{AA}=\frac{1}{1+2\,C\la r_T^2\ra\,T_A}\ 
\Leftrightarrow\  \frac{1}{\bigl[1+C\la r_T^2\ra\,T_A\bigr]^2}
\label{260}
\eeq
One can see that the two absorption factors are quite different, especially for $C\la r_T^2\ra T_A\gsim1$.
The source of the difference is color filtering. Namely, the mean transverse size of a $\bar cc$ wave packet propagating through a nucleus is getting smaller, since large-size dipoles are filtered out (absorbed) with a larger probability. Such a dipole with a reduced mean size easier penetrates through
the second colliding nucleus, compared to what would be in $pA$ collision. The mutual color filtering makes both nuclei more transparent.

\begin{figure}[htb]
\begin{center}
\includegraphics[width=6cm]{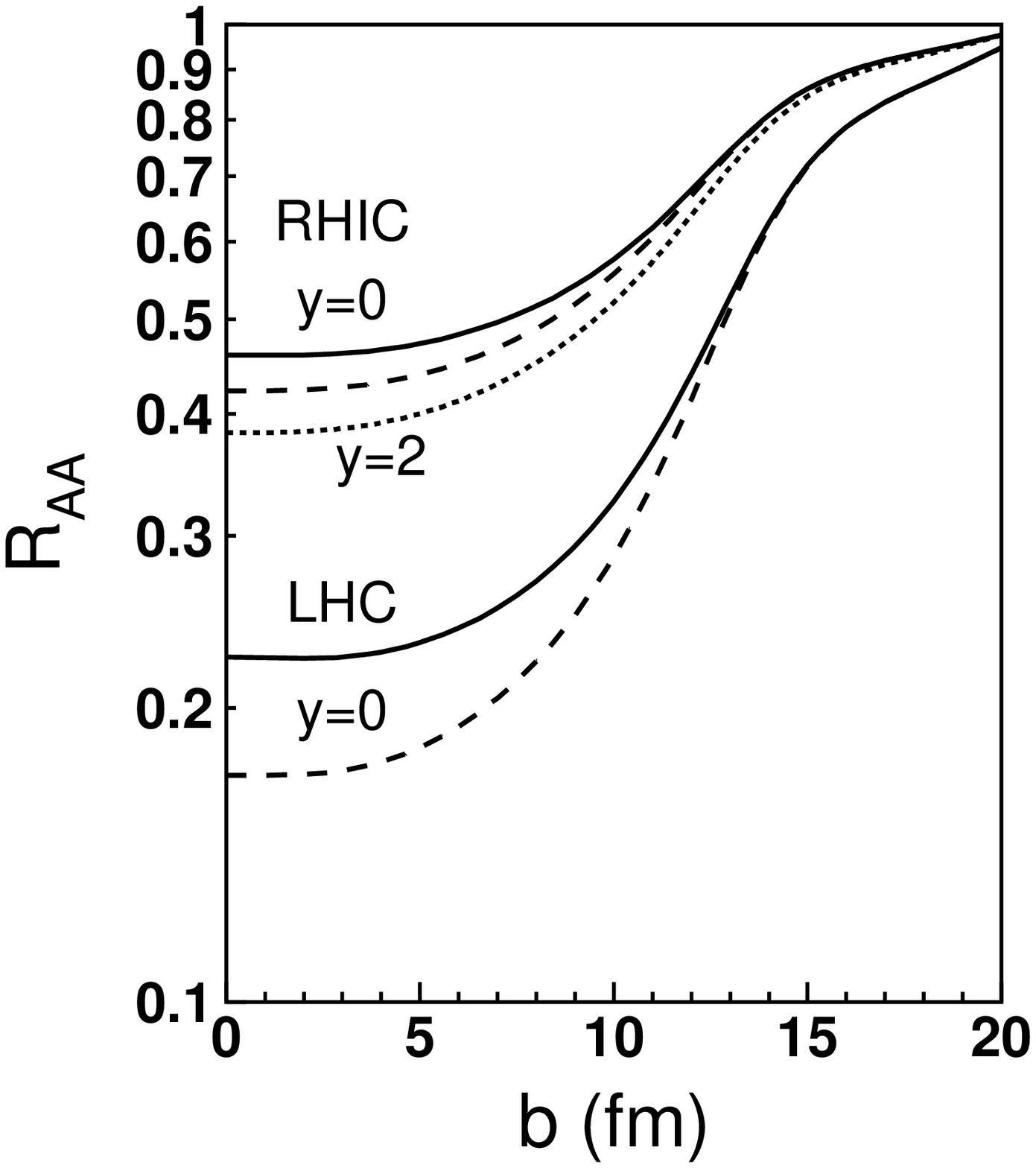}
\hspace{10mm}
\includegraphics[width=6cm]{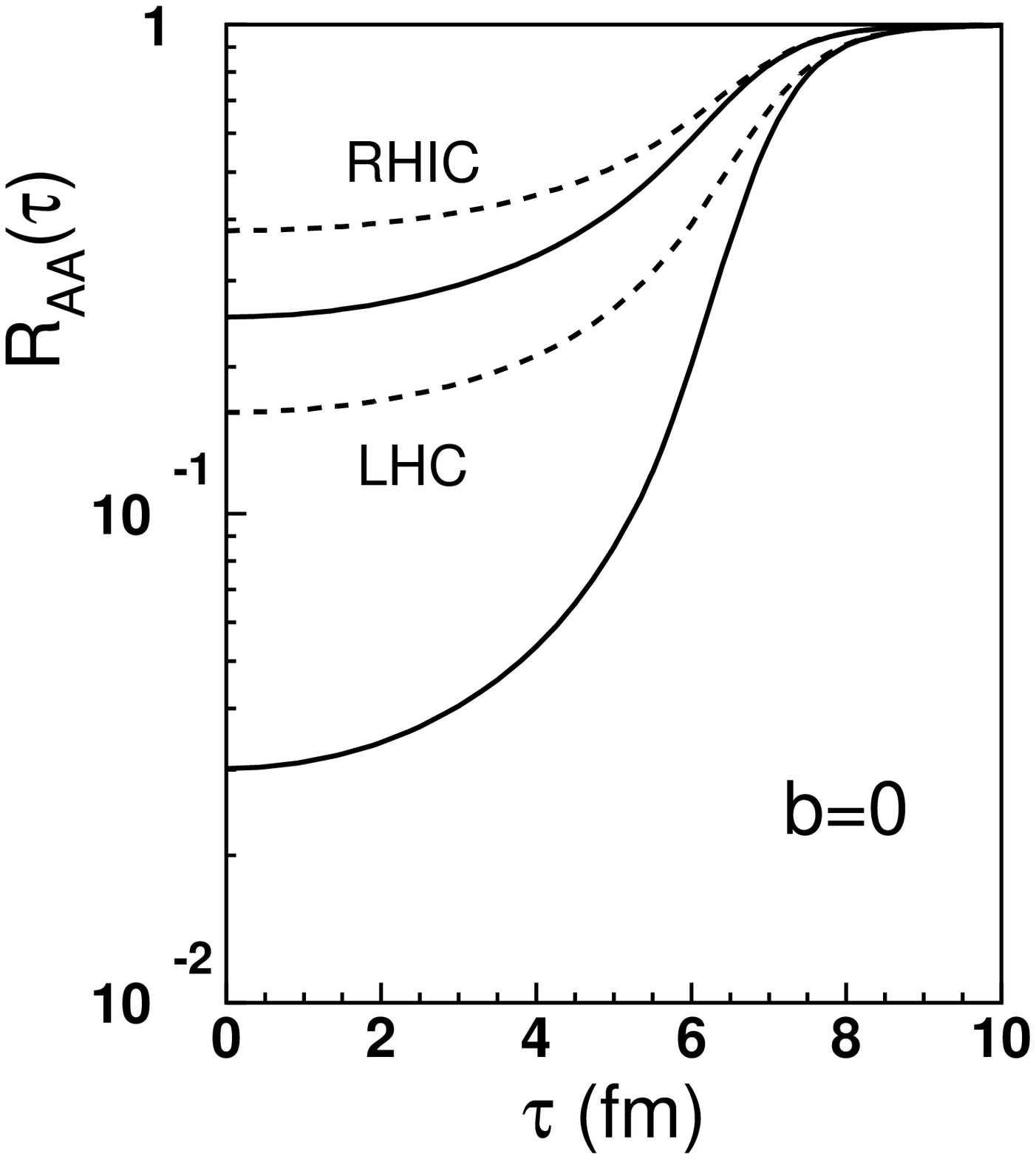}
\end{center}
\caption{\label{A-A} {\it Left:} Effects of double-color-filtering. $\J$ suppression by the initial state interaction (ISI) effects in gold-gold collisions at $\sqrt{s}=200\GeV$ as function of $b$. The upper and bottom pairs of curves (solid or dashed) correspond to $y=0$ and energies $\sqrt{s}=200\GeV$ and $5.5\TeV$ respectively. 
Solid and dashed curves present the results at $y=0$ including  and excluding the effect of double-color-filtering, respectively. The dotted curve demonstrates rapidity dependence of the ISI effects at RHIC. It is calculated at $y=2$ and is to be compared with the upper solid curve at $y=0$.
{\it Right:} Effects of boosted saturation scale. The upper and bottom dashed curves correspond
to central gold-gold collisions at the energies $\sqrt{s}=200\GeV,\ 5.5\TeV$ respectively. They demonstrate dependence on impact parameter $\tau$ and are calculated in the same way as the solid curves in the left panel. Solid curves are calculated with the boosted saturation scale, which makes the nuclei more opaque for heavy dipoles.
}
 \end{figure}

Now we are in a position to perform realistic calculations for the nuclear suppression factor in $AB$ collisions. Provided that the $\bar cc$ production occurs in the long coherence length regime for both nuclei, the nuclear suppression factor at impact parameter $b$ reads,
\beqn
R_{AB}(b)&=&{1\over T_{AB}(b)}\int d^2\tau\,\frac{T_A(\tau)T_B((\vec b-\vec\tau)}
{(\Lambda_A^+-\Lambda_A^-)(\Lambda_B^+-\Lambda_B^-)}
\nonumber\\&\times&
\ln\left[\frac{(1+\Lambda_A^-+\Lambda_B^+)(1+\Lambda_A^++\Lambda_B^-)}
{(1+\Lambda_A^++\Lambda_B^+)(1+\Lambda_A^-+\Lambda_B^-)}\right]
\label{280}
\eeqn
where
\beqn
\Lambda_{A(B)}^+&=&{\la r_T^2\ra\over2}\,C(E^{A(B)}_{\bar cc})T_{A(B)};
\label{300a}\\
\Lambda_{A(B)}^-&=&{7\la r_T^2\ra\over32}\,C(E^{A(B)}_{\bar cc})T_{A(B)},
\label{300b}
\eeqn
and $E^{A,B}_{\bar cc}$ are the energies of the $\bar cc$ in the rest frames of the nuclei $A$ and $B$ respectively. The result of calculation of Eq.~(\ref{280}) is plotted by the upper solid curve in the left panel of Fig.~\ref{A-A}. 
For comparison, the result of conventional calculations assuming simple multiplication of the suppression factors in the two nuclei, is depicted by dashed curve. We see that the mutual color filtering makes the nuclei considerably more transparent. This effect should be more prominent for production of $\Psi'$ and $\chi$. 

With Eq.~(\ref{280}) we can trace the $y$-dependence of $R_AA$. It turns out to be rather weak
at the energy of RHIC, what obviously follows from the approximate linearity of $y$-dependence in $pA$ depicted in the left panel of Fig.~\ref{d-A}. However, at sufficiently large $y$, say $y=2$,
the condition of long coherence length breaks down in one of the nuclei. Then the $\bar cc$ dipole size is not frozen by Lorentz time delation, and the filtering in this particular nucleus is not effective any more. In this case the conventional multiplicative procedure is applicable, but the suppression factor in one nucleus (high $E_{\bar cc}$) should be calculated according to Eq.~(\ref{160}), while in another nucleus (low $E_{\bar cc}$) one should do calculations for the short $l_c$ regime with the $L$-dependent absorption cross section Eq.~(\ref{80}). The result of such calculation is plotted by the bottom solid curve in the right panel of Fig.~\ref{d-A}. We see that the nuclear suppression at $y=2$ is stronger than at $y=0$. This happens due to disappearance of the double-color-filtering effect.

\subsection{Boosted saturation scale in $AA$ collisions}

Another mechanism which breaks down the conventional multiplicative procedure for the transition from $pA$ to $AA$ is the mutual boosting of the saturation scale in $AA$ collisions compared with $pA$ \cite{boosting}. It significantly increases the break-up cross section up to factor 1.5 and factor 3 at the energies of RHIC and LHC respectively. Correspondingly, the nuclear medium turns out to be much more opaque for $\bar cc$ dipoles in the case of nuclear collisions compared with the simplified multiplicative prescription of \cite{phenix-dA,cassagnac}.

In the right panel of Fig.~\ref{A-A} we demonstrate  the strength of this effect for central ($b=0$) gold-gold collision as function of impact parameter $\tau$. The upper and bottom dashed curves corresponding to the energies of RHIC and LHC respectively, include the double-color-filtering effect, but exclude the saturation scale boosting, which is added to produce the solid curves.
 
Thus, $\J$ should be suppressed in $AA$ collisions significantly stronger, than usually expected, and one should not interpret that as an anomalous suppression caused by final state interaction (FSI) with the dense medium. 

\section{Propagation of $\J$ through a dense medium}

In the c.m. of nuclear collision the nuclear disks passing through each other leave behind a cloud of radiated gluons creating a dense matter, which the $\J$ propagates through. In this reference frame the $\J$ full momentum is $p_T$, which ranges from zero to several GeV in RHIC data.
Such a low energy  $\bar cc$ dipole develops the $\J$ wave function pretty fast, during time $t_f<0.5\fm$ \cite{kps-psi}, which is about the time scale of the medium creation.
Thus, what is propagating through the medium is not a small $\bar cc$ dipole, but a fully formed $\J$. 
The mean dipole cross section is $\sigma_{\J}={2\over3}C\,\la r^2_{\J}\ra$, where the factor $C$ was introduced in (\ref{80}) and is known for a proton target 
from DIS data \cite{hikt}. Its value for a hot medium is unknown, however, the factor C also controls broadening of a quark propagating through the medium \cite{jkt}. So it is related to the transport coefficient $\hat q$ \cite{bdmps}, which is in-medium broadening per unit of length, 
$C=\hat q/2\rho$. 

Therefore, the survival probability of $\J$ produced at impact parameter $\vec\tau$ inside the medium has the form,
\beq
R_{AA}^{FSI}(\vec\tau,p_T)\Bigr|_{b=0}=
\int\limits_0^\pi \frac{d\phi}{\pi}
\exp\Biggl[-{1\over3}\la r_{\J}^2\ra\int\limits_{l_0}^\infty dl\,
\hat q(\vec\tau+\vec l)
\Biggr],
\label{140a}
\eeq
where $|\vec\tau+\vec l|^2=\tau^2+l^2+2\tau l\cos\phi$; $l_0=vt_0$; and $t_0=0.5\fm$.

The transport coefficient depends on the medium density, which is function of impact parameter and time. We rely on the conventional form \cite{frankfurt},
\beq
\hat q(t,\vec b,\vec\tau)=\frac{\hat q_0\,t_0}{t}\,
\frac{n_{part}(\vec b,\vec\tau)}{n_{part}(0,0)},
\label{520}
\eeq
where $\vec b$ and $\vec \tau$ are the impact parameter of the collision and of the point where the $\hat q$ is defined, respectively. The transport coefficient $\hat q_0$ corresponds to the maximal medium density produced at impact parameter $\tau=0$ in central gold-gold collision at the time $t=t_0$ after the collision. We treat $\hat q_0$ as a adjusted parameter. 

\begin{figure}[htb]
\begin{center}
\includegraphics[width=6cm]{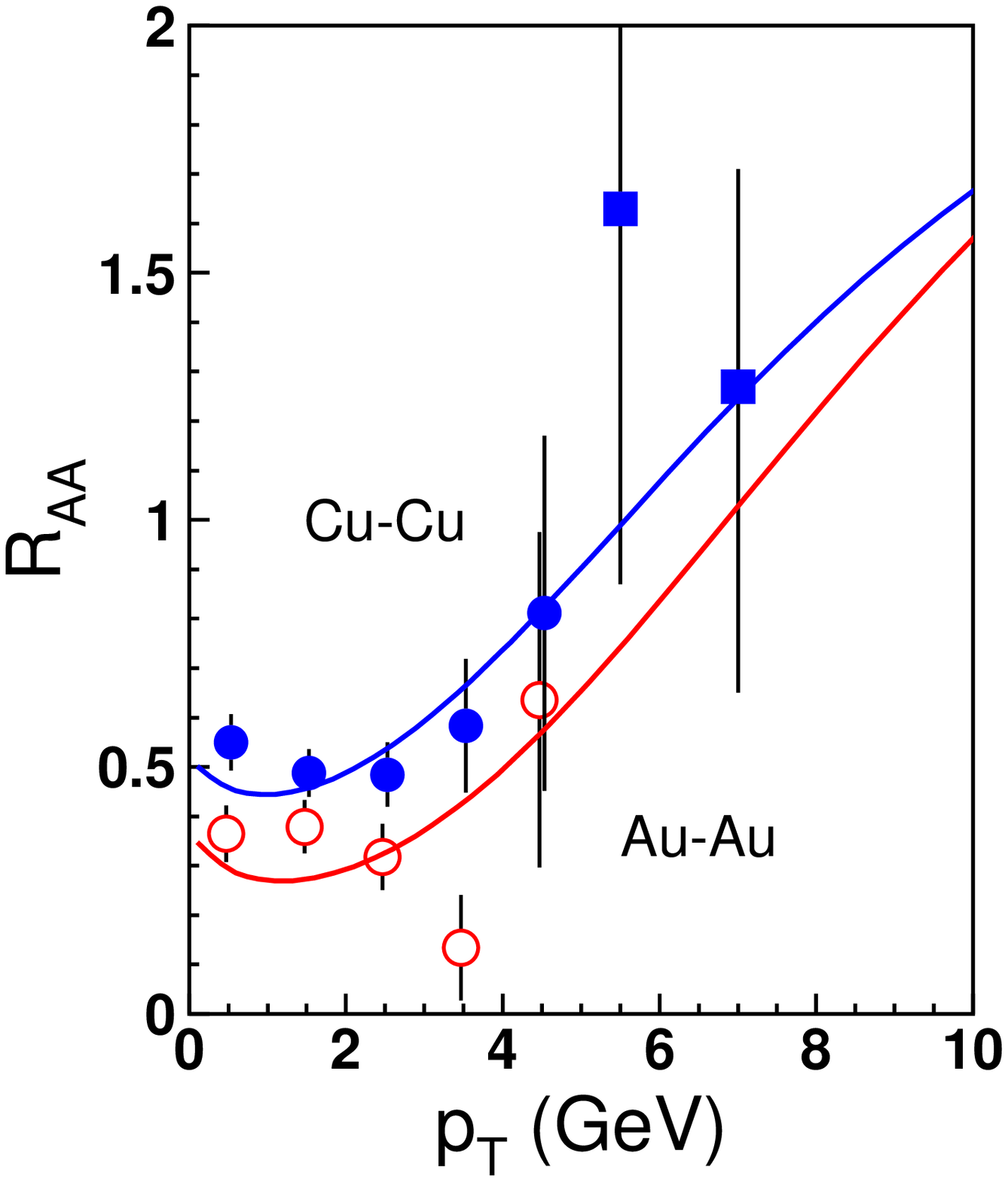}
\hspace{10mm}
\includegraphics[width=6cm]{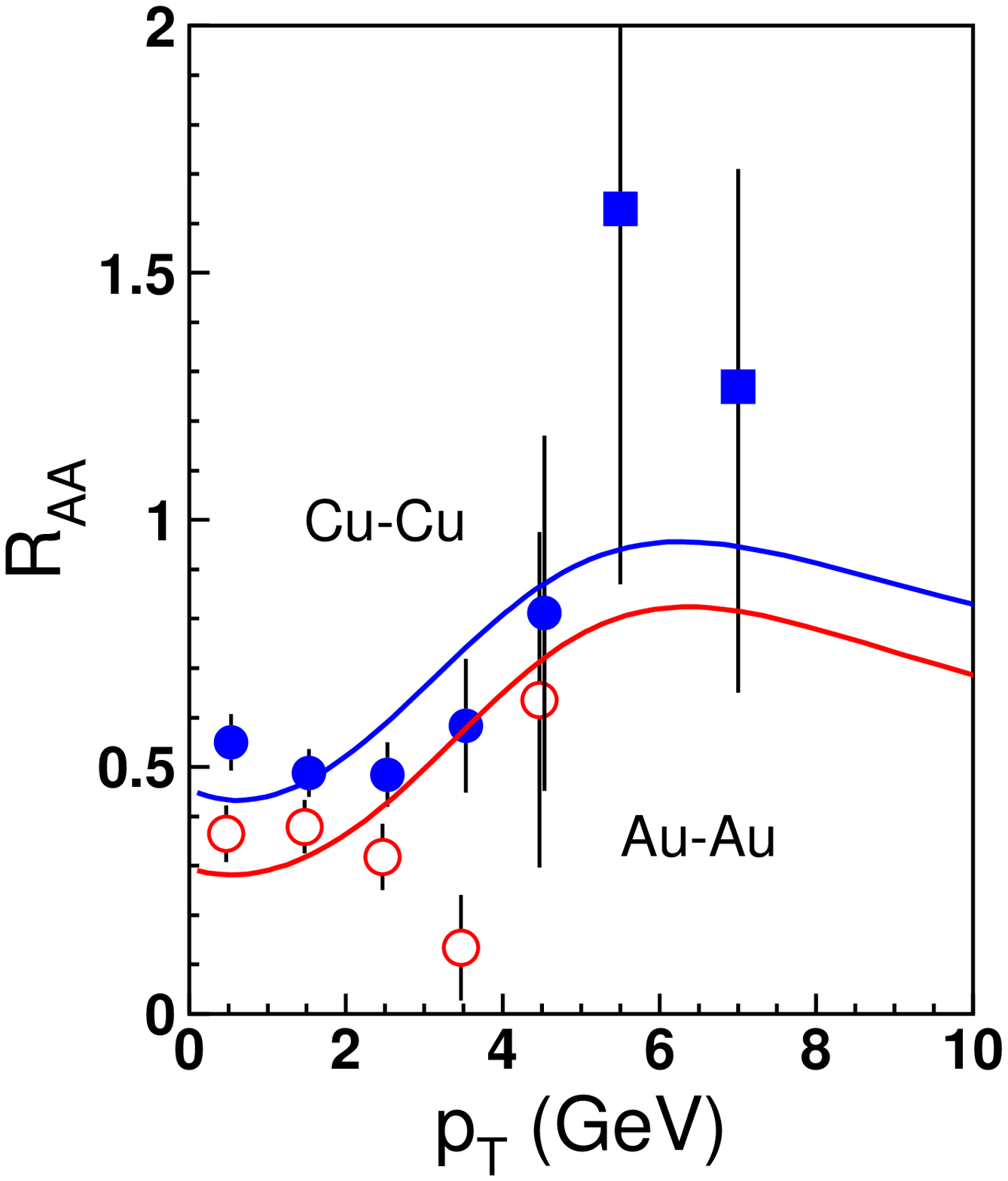}
\end{center}
\caption{\label{Cu+Au}  Nuclear ratio $R_{AA}$ for central copper-copper (full circles and squares, upper curve) and gold-gold (empty circles, bottom curve) collisions at $\sqrt{s}=200\GeV$ as function of $\J$ transverse momentum. The curves in the left and right panels differ by calculation of the Cronin effect as described in text.
Data are from \cite{phenix-psi,star-psi}.}
 \end{figure}

The observed nuclear effects in $\J$ production in $AA$ collisions is interpreted as a combination of FSI of $\J$ in the dense medium Eq.(\ref{140a}), and the initial state interaction (ISI) effects in production of
$\J$ caused by multiple interactions of the colliding nuclei. The latter was discussed above and includes attenuation of the produced $\bar cc$ dipole propagating through both nuclei, high twist shadowing of charm quarks, and leading twist gluon shadowing.
In addition, gluon saturation in nuclei \cite{broad} leads to a considerable broadening of gluons, which causes a strong Cronin effect for $\J$. The details of calculations of the ISI effects can be found in \cite{kps-psi}.
The results for the nuclear effects in copper-copper and gold-gold are presented in Fig.~\ref{Cu+Au}.
The two plots differ by calculations of the Cronin effect. One can either make a shift in $\la p_T^2\ra$
caused by broadening (left), or make a convolution of a nuclear-modified primordial 
transverse momentum distribution of the colliding partons with the known $p_T$-distribution of $\J$ (right). Both results are similar, except at large $p_T>5\GeV$, where no $pp$ data are available, therefore no reliable prediction can be made.

All effects are evaluated in a parameter free way, except the transport coefficient, which should be in the range of  $\hat q_0\approx 0.3-0.5\GeV^2/\fm$ in order to reproduce data. This is close to the expected value $\hat q_0\approx0.5\GeV^2/\fm$ \cite{bdmps}, and more than order of magnitude less than was found from jet quenching data within the energy loss scenario \cite{phenix-theory}.
Notice that the cold nuclear matter ISI suppression might have been underestimated in \cite{kps-psi}.
Our current parameter-free estimate made in Sect.~\ref{double-filtering} results a significantly stronger ISI suppression, which is almost sufficient to explain RHIC data on $\J$ production in central gold-gold collisions. This means that $q_0$ may be even smaller, challenging the claim that a dense matter is created.

\section{Summary}

This talk highlighted several unusual features of $\J$ production in $pA$ and $AA$ collisions,
currently  debated in the literature, which can be understood taking a deeper look at the underlying dynamics.
Since the wide spread interpretation of $\J$ production off nuclei is grossly oversimplified, some improvements are proposed.

\begin{itemize}

\item 
At the energies of fixed target experiments at SPS and Fermilab the break-up cross section for a $\bar cc$ dipole is subject to color transparency and is fluctuating  during propagation through the nucleus.
A simple model for the break-up cross section, which depends on energy and path length is developed.
It well explains the energy dependence of the effective absorption cross section observed in data, and its absolute value.

\item
At high energies of RHIC and LHC the charm production time becomes long, leading to a higher twist shadowing. This effect is of the same order as  the attenuation caused by the $\bar cc$ break-up, and the magnitude of both is well fixed by DIS data from HERA. 
Data for $\J$ suppression in $dA$ collisions at RHIC are well explained without adjustment.

\item
On the contrary, leading twist gluon shadowing is found to give no contribution to available RHIC data
for $\J$ suppression, and to be a rather small correction even at the energies of LHC.

\item
Another effect, which makes transition from $pA$ to $AA$ model dependent is double color filtering.
When a $\bar cc$ dipole propagates simultaneously through the colliding nuclei, and one nucleus filters out large size dipoles, the reduced mean dipole size makes another nucleus more transparent.

\item
Multiple interactions in the colliding nuclei lead to involvement of higher Fock states in the bound nucleons, which in turn enhance the multiple interactions and bust the saturation scale further up.
As a result, the nuclear medium becomes significantly more opaque for $\J$ in $AA$ compared with $pA$ collisions.

\item
RHIC data for $\J$ suppression in $AA$ collision is well described combining the above effects
and fitting the density of the produced hot medium, which is characterized by a transport coefficient $\hat q_0$. The found value of $\hat q_0$ is in good agreement with theoretical expectations, and is substantially smaller than what was extracted from jet quenching data. Thus, $\J$ production can serve as an efficient probe for the density of the created matter.

\end{itemize}

\section*{Acknowledgments}

I am thankful to my collaborators Hans-J\"urgen Pirner, Irina Potashnikova, Ivan Schmidt and Sasha Tarasov for help with calculations and numerous discussions. This work was supported in part
by Fondecyt (Chile) grant 1090291, by DFG
(Germany) grant PI182/3-1, and by Conicyt-DFG grant No. 084-2009.

\bibliographystyle{elsarticle-num}

\end{document}